\documentclass[twocolumn,prl,preprintnumbers,amsmath,amssymb]{revtex4-1}
\usepackage{graphicx}
\usepackage{dcolumn}
\usepackage{bm}
\usepackage{tabu}

\begin{document}
\title{Homogeneous nucleation of ice}

\author{Haiyang Niu$^{1,2}$}

\email[]{haiyang.niu@phys.chem.ethz.ch}

\author{Yi Isaac Yang$^{1,2}$}

\author{Michele Parrinello$^{1,2,3}$}
\email[]{parrinello@phys.chem.ethz.ch}

\affiliation{$^1$ Department of Chemistry and Applied Biosciences, ETH Zurich c/o USI Campus, Via Giuseppe Buffi 13, 6900 Lugano, Switzerland}
\affiliation{$^2$ Facolt\`{a} di Informatica, Instituto di Scienze Computationali,
  and National Center for Computational Design and Discovery of Novel Materials MARVEL,
  Universit\`{a} della Svizzera italiana (USI), Via Giuseppe Buffi 13, 6900 Lugano, Switzerland}
\affiliation{$^3$ Istituto Italiano di Tecnologia, Via Morego 30, 16163 Genova, Italy}

\date{\today}
\begin{abstract}

Ice nucleation is a process of great relevance in physics, chemistry, technology and environmental sciences, much theoretical and experimental efforts have been devoted to its understanding, but still it remains a topic of intense research. We shed light on this phenomenon by performing atomistic based simulations. Using metadynamics and a carefully designed set of collective variables, reversible transitions between water and ice are able to be simulated. We find that water freezes into a stacking disordered structure with the all-atom TIP4P/Ice model, and the features of the critical nucleus of nucleation at the microscopic level are revealed. Our results are in agreement with recent experimental and other theoretical works and confirm that nucleation is preceded by a large increase in tetrahedrally coordinated water molecules. 

\end{abstract}


\maketitle

Ice nucleation from water is an ubiquitous phenomenon~\cite{bartels2012}, relevant in many areas of science and technology, from atmospheric and environmental science, to aviation technology and to biology. Understanding microscopically this process is of great value~\cite{bartels2013,coluzza2017,kiselev2017}.
Interest in homogeneous ice nucleation in undercooled water stems also the fact that it occurs in the so called no man's land temperature region. Thus it offers a tool to investigate the behavior of water in this part of the phase diagram that is deemed to be important for the understanding of water anomalies. Unfortunately a direct simulation of this phenomenon is not possible given the time scale over which crystallization takes place. This has made difficult to reproduce the pioneering work of Ohmine and coworkers\cite{matsumoto2002}, in which one spontaneous nucleation event was reported. For this reason a number of simulations with different methods, i.e., seeding approaches \cite{sanz2013, espinosa2016,espinosa20162,lupi2017,cheng2018}, enhanced sampling methods~\cite{lupi2017,quigley2008,reinhardt2012a,reinhardt2012b,geiger2013,pipolo2017}  or forward-flux sampling~\cite{li2011,li2013,haji2015}, have been carried out. However, reversible transitions between ice and water and direct nucleation of ice remain a great challenge, especially when using an all-atom water potential. 

In this Letter we carry out metadynamics~\cite{laio2002,barducci2008} simulations to investigate ice nucleation, a well tested and much used method whose validity has been rigorously established~\cite{Laio2008}. In common with other enhanced sampling methods~\cite{torrie1977,valsson2014} metadynamics requires the introduction of appropriate collective variables (CVs). If the CVs are properly chosen, that is if they reflect the physics underlying the process, convergence is smooth. In a different context, we have shown that the X-ray diffraction (XRD) peak intensities are useful CVs in the study of crystallization. For instance they have been successfully applied to a system as complex as silica~\cite{niu2018}. In some way silica has a behavior not too dissimilar from water. For instance it exhibits a density anomaly~\cite{shell2002}. Thus it felt natural to continue using the same class of CVs. 

Here we propose two collective variables, one is a suitable combination of scattering peak intensities and has a long range character. The other is a surrogate for translational entropy that has a more local nature. Importantly, these collective variables do not prejudice the ice structure to be formed. Using the all-atom TIP4P/Ice model~\cite{abascal2005} we find that water freezes into a stacking disordered structure. We can follow in detail the nucleation process and the features of the critical nucleation nucleus have been discussed. 

Even though hexagonal ice (ice $I_h$) is the stable crystal phase at ambient pressure, stacking-disordered ice (ice $I_{sd}$) consisting of random sequences of cubic and hexagonal ice layers are commonly observed both in experiments and simulations~\cite{murray2005,malkin2012,haji2015,lupi2017,pipolo2017,moore2011b}. We chose a CV that is blind with respect to the form of ice polytypes that can be formed, $I_h$, $I_{sd}$, or cubic ice $I_c$. To this effect one of the CVs is constructed as a linear combination of seven descriptors ~\cite{online} (See the Supplemental Material~\cite{online} and Fig. S1 for how we choose these descriptors):
\begin{equation}\label{E:sx}
  s_X= s_{100} + s_{002} + s_{101} + \alpha(s^{xy}_{100} +s^{xy}_{\bar{1}20} ) + \beta s^{xz}_{002} +\gamma x^{yz}_{002} \,,
\end{equation}
in which the first three descriptors correspond to the X-ray diffraction intensities of the three main peaks of the system, and $s^{xy}_{100}$ and $s^{xy}_{\bar{1}20}$ correspond to the intensities of the two main peaks of one single honeycomb bi-layer which is projected into the $x$-$y$ plane, the last two descriptors $s^{xz}_{002}$ and $s^{yz}_{002}$ refer to the intensity of the first main peak of the layers which are vertical to the honeycomb bi-layer in the $x$-$z$ and $y$-$z$ planes. The coefficients $\alpha$, $\beta$ and $\gamma$ adjust the weights between different descriptors, here in this work $\alpha$ = 2, $\beta$ = 1 and $\gamma$ = 1. This particular combination has proven to be efficient in accelerating nucleation.

\begin{figure}
  \centering
  \includegraphics[width=0.95\columnwidth]{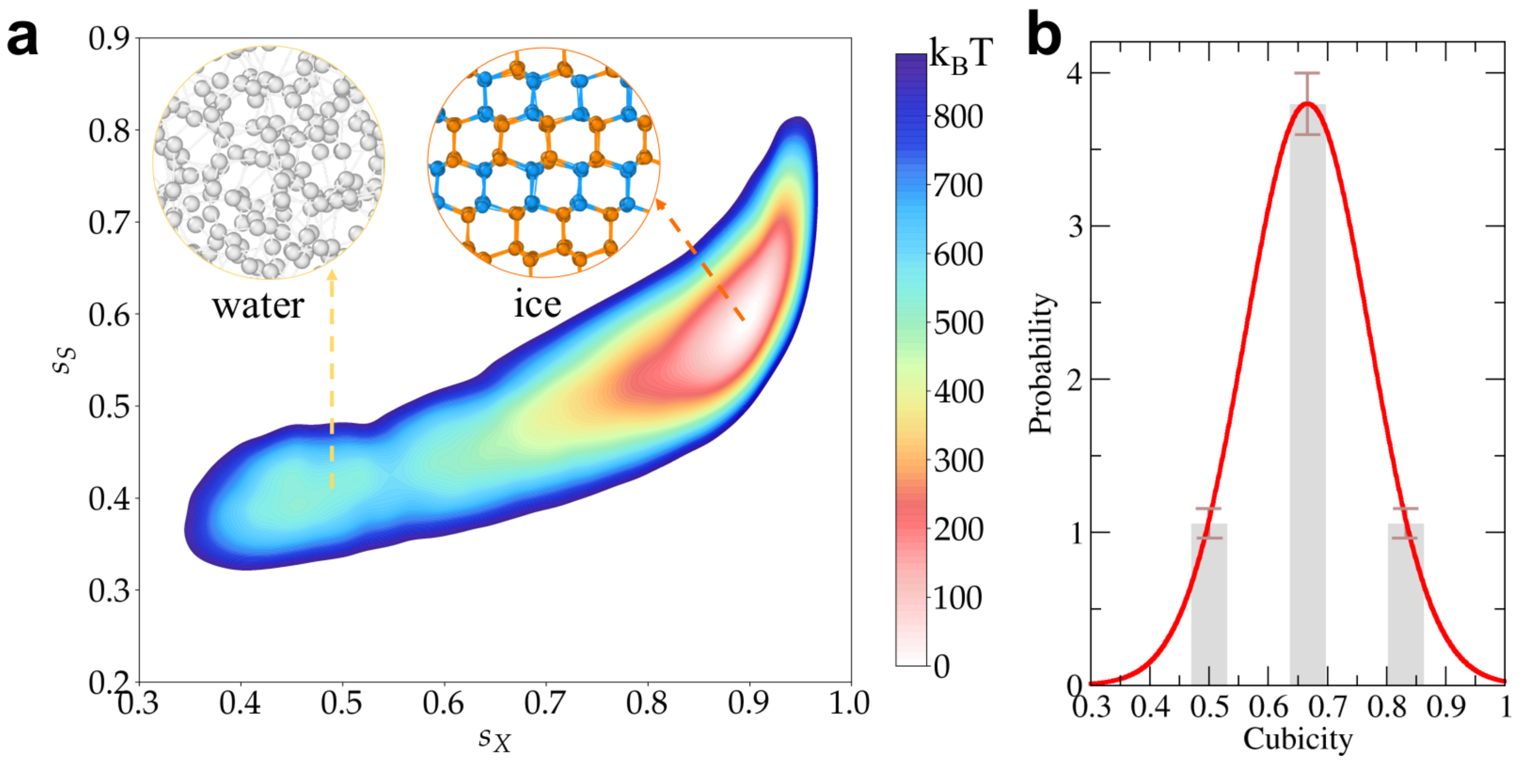}
  \caption{ Free energy surface in terms of collective variables $s_{X}$ and $s_S$ with TIP4P/Ice model. a) Free energy surface at 230K for a system of 1600 water molecules. b) Probability distribution of the cubicity of the nucleated ice structure obtained from simulations.}
  \label{F:cv}
\end{figure}

In addition we found useful to combine the long-range order CV $s_X$ with another CV that has a more local character, that is the surrogate for translational entropy $s_S$ which has been successfully used elsewhere~\cite{piaggi2017,piaggi2018}. This additional CV proved important in accelerating the simulation convergence. The simulation that use $s_X$ and $s_S$ converge much faster than those using only $s_X$ in spite of the higher dimensionality of the CV space (see Fig. S2)~\cite{online}. This could be ascribed to the fact that $s_S$ accelerates melting of ice and during crystallization helps clearing defects. One the other hand, long-range order based CVs are essential in simulating ice nucleation, since $s_S$ alone cannot lead to nucleation.

Full technical details can be found in the Supplemental Material~\cite{online}. Here we only recall that we have used the TIP4P/Ice model of water~\cite{abascal2005}, that has been explicitly designed to describe the solid phases of water. Our simulation were performed at 230 K, close to the experimental homogeneous ice nucleation temperature. The pressure has been set to its atmospheric value using a Parrinello-Rahman barostat~\cite{parrinello1981} that allows only orthorhombic fluctuations. 

In Fig. S2-3~\cite{online} we see how the two order parameters allow to go reversibly from liquid to a stacking disordered ice $I_{sd}$, in consistent with experimental and other theoretical works~\cite{murray2005,malkin2012,haji2015,lupi2017,pipolo2017}. The reversible transitions allow us to draw in Fig.~\ref{F:cv}a the free energy surface(FES) as a function the chosen CVs. As expected, at this temperature the solid phase minimum is much lower than the liquid one. On this surface an almost linear free energy path can be tracked that goes from liquid to solid. The barrier to this transition is $\bigtriangleup$G = 52.8 $\pm$6 $k_BT$, in agreement with other theoretical predication based on classical nucleation theory~\cite{haji2015}. To quantify the characteristic of the stacking disordered ice $I_{sd}$ that obtained from nucleation, we have calculated the cubicity, i.e. the fraction of cubic stacking sequences, of the obtained solids for over 60 nucleation events, the results shows the probability follows a distribution as shown in Fig.~\ref{F:cv}b, with mean cubicity value $C^*$ = 0.67 $\pm$ 0.10, comparably to that ($C^*$ = 0.63$ \pm $0.05) obtained with the coarse-grained MW potential~\cite{lupi2017}. Furthermore, the cubic and hexagonal sequences are randomly arranged, which result into a lower symmetry trigonal $I_{sd} $ structure with the space group of $P3m1$(See Fig. S4)~\cite{malkin2015,online}.

\begin{figure}
  \centering
  \includegraphics[width=0.95\columnwidth]{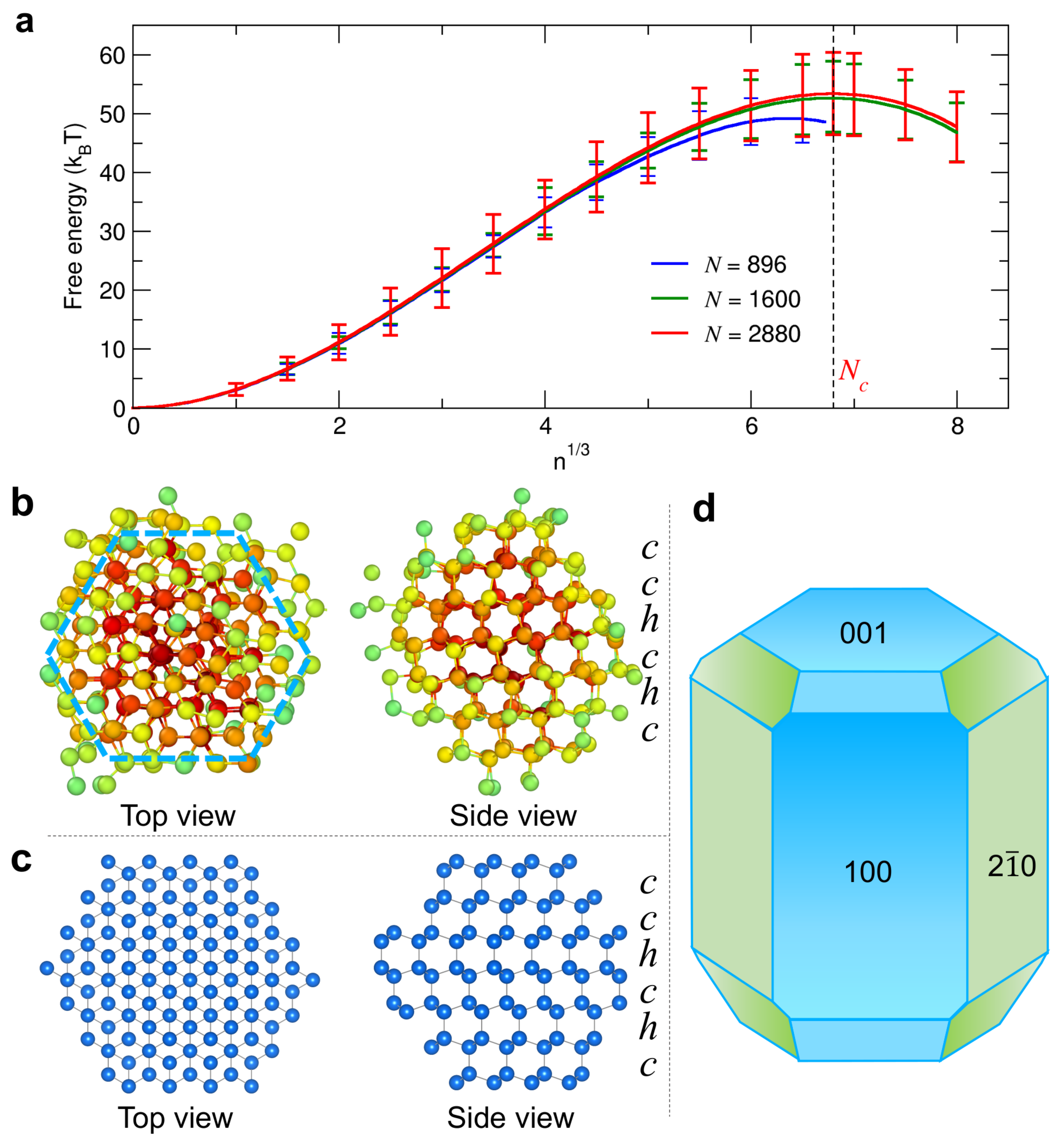}
  \caption{Features of critical nucleus of ice nucleation. a) Reweighted free energy as a function of ice cluster size $n^{1/3}$ for system with N = 896, 1600 and 2880 water molecules. b-c) A typical critical ice nucleus and an idealized crystalline model of it. Here \textit{h} and \textit{c} refer to the hexagonal and cubic sequences of $I_{sd}$. d) Possible crystal shape of $I_{sd}$ with a threefold rotational symmetry.}
  \label{F:nucleation}
\end{figure}

Having harvested a large number of crystallization events we have enough statistic to address the issue of nucleation.
In classical nucleation theory (CNT), that is the theoretical cornerstone in nucleation studies, the free energy as a function of the solid-like cluster size plays a pivotal role. For this reason we first identify a variable that is able to distinguish between solid-like and liquid-like atoms. In the spirit of our work we sit on each atom and calculate the instantaneous scattering intensity of that particular atom to the CV $s_X$ (See Fig. S5)~\cite{bonati2018,online}. This fingerprint is able to distinguish well between solid-like and liquid-like atoms. Then we identify all the solid-like atom cluster and make a histogram as a function of n$^{1/3}$ where n is the number of atom in a solid-like cluster. The quantity n$^{1/3}$ is proportional to the cluster radius. In order to determine system size effects, we have performed calculations for three different systems with N = 896, 1600 and 2880 water molecules respectively. Fig.~\ref{F:nucleation}a shows that the curve of N = 1600 is indistinguishable with that of N = 2880, while divergence can be noticed for that of N = 896 at relative higher $n$. Thus system size effects can be ruled out for system sizes larger than 1600 since from this size on the critical nucleus fits into the simulation box. Our estimate of the critical nucleus (see Fig.~\ref{F:nucleation}b) size gives a value of $N_c$ = 314 $\pm$ 20, in agreement with other theoretical estimates~\cite{haji2015}. This number is tempting close to $N_c$ = 321 that is the number of water molecules contained in a microcrystallite whose shape is depicted in Fig.~\ref{F:nucleation}c. This suggestion is strengthened by the fact that some faceting can be observed in a visual inspection. We must add that in the morphology of the typical critical nucleus, variations around this shape are seen. The microcrystallite in Fig.~\ref{F:nucleation}c thus has to be thought of as an idealized representation of the critical nucleus shape. According to this picture the critical nucleus is close to to being spherical and appears to have a threefold rotational symmetry in accordance with the space group($P3m1$) of $I_{sd}$, which can be expected to grown into a trigonal symmetry ice (Fig.~\ref{F:nucleation}d)~\cite{murray2015}. 
In order to check independently whether this nucleus belongs to the transition state ensemble, we have performed as many as 50 independent trajectories starting from a water configuration that has been equilibrated in the presence of the ideal crystalline in Fig.~\ref{F:nucleation}c. In the equilibiration time the atomic positions of the crystalline were held into place by a restraining potential. Once this potential was released, unbiased molecular dynamics simulations were performed. In 50 independent simulations, around 56$\%$ and 44$\%$ trajectories show that the ice nuclei grow and melt, respectively, which indicates our estimation of the critical size is in the right ballpark. Furthermore, some trajectories show that the ice nuclei neither grow nor melt in 500 ns simulations, which indicates that the potential energy surface around the critical ice cluster is rather flat, coherent with the transition state shape in Fig.~\ref{F:nucleation}.

\begin{figure}
  \centering
  \includegraphics[width=0.95\columnwidth]{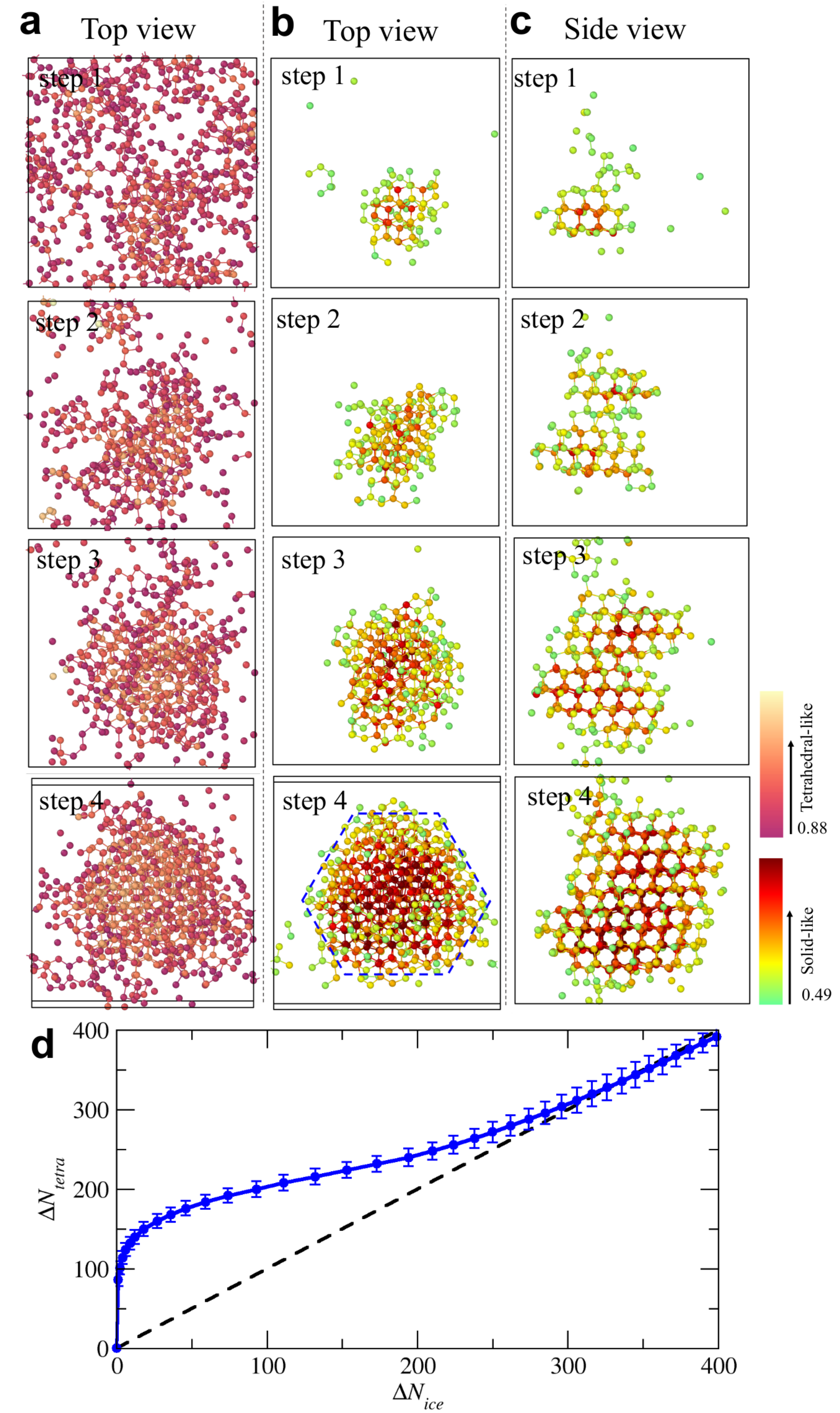}
  \caption{ Homogeneous ice nucleation process.  Configuration snapshots at different stages of nucleation of tetrahedral-like atoms (a), and solid-like atoms (b-c) are shown, respectively. d) Relationship between the number of solid-like atoms $\Delta$$N_{ice}$ and tetrahedral-like atoms $\Delta$$N_{tetra}$. Here $\Delta$$N_{tetra}$ refers to the relative tetrahedral-like atom number to that of liquid state. The dashed line is drawn only to guide the eyes.}
\label{F:process}
\end{figure}

Much theoretical and experimental works~\cite{sellberg2014,tanaka2012,moore2011a,bullock2013,Fitzner2018,errington2002,russo2014,overduin2015,pirzadeh2011} have been devoted to understand the mechanistic insight of the structural transformation from water to ice. With the coarse-grained MW potential, Molinero and coworkers ~\cite{moore2011a,bullock2013} have uncovered that the rate of homogeneous nucleation of ice is controlled by the structural transformation into a four-coordinated liquid, and ice nucleates mostly within the four-coordinated liquid patches. In order to analyze the nucleation process with the all-atom TIP4P/Ice model, we have plotted the configuration snapshots of one typical nucleation process (Fig.~\ref{F:process}), in which the tetrahedral-like and ice-like atoms are tracked. The tetrahedral-like atom is identified by the tetrahedral order parameter~\cite{chau1998,errington2001}, in which both the distances and angles between the oxygen atoms are taken into account. Our results show that the patches of tetrahedrally coordinated water molecules play a central role as precursor structure for ice nucleation and ice clusters nucleate from these patches in agreement with experimental~\cite{sellberg2014} and other theoretical works~\cite{tanaka2012,moore2011a,bullock2013,overduin2015}. This can be seen in Fig.~\ref{F:process}d that before the critical nucleus size is reached the number of tetrahedral-like water $\Delta$$N_{tetra}$ is much larger than those that have a local solid-like environment, which is in accordance with Moore and Molinero observations with the coarse-grained MW potential~\cite{moore2011a}. As the nucleation process $\Delta$$N_{ice}$ - $\Delta$$N_{tetra}$ becomes smaller until $\Delta$$N_{ice}$ $\approx$ $\Delta$ $N_{tetra}$ in the proximity of the critical size. These results can also explains why the CV $s_S$ with local structure nature is essential to enhance ice nucleation. In our simulations, we also observed that several ice clusters can be formed in the simulation box, in some cases one cluster grows while others dissolve, in others two or even more clusters can grow together until merging into one cluster(See Fig. S6)~\cite{online}. The size of such merged cluster can instantaneously surpass the size of the critical cluster.

In this Letter, our findings show that the formation of ice from undercooled water can be successfully simulated with the all-atom TIP4P/Ice model by enhanced sampling method. To induce this transition, we have proposed two collective variables. They are the intensity of properly selected scattering peaks that are more sensitive to long range order and a surrogate for translational entropy that gives local information. Our results demonstrate that stacking disordered ice can be formed directly from water at homogeneous conditions with a mean cubicity $C*$ = 0.67 $\pm$ 0.10 and critical size $N_c$ = 314  $\pm$ 20 at 230 K. The barrier of ice nucleation at this temperature is estimated to be $\bigtriangleup$G = 52.8 $\pm$6 $k_BT$. We also find that ice nucleates from the tetrahedrally coordinated structure patches in agreement with experimental and other theoretical works, and the ice embryo grows with threefold rotational symmetry. Our work makes the ice formation study with all-atom water potential possible, and is a starting point for more sophisticated ice nucleation problem, such uncovering active sites in heterogeneous ice nucleation and anti-freezing protein study.

\begin{acknowledgments}
This research was supported by the NCCR MARVEL, funded by the Swiss National Science Foundation, and the European Union Grant No. ERC-2014-AdG-670227/VARMET.
  The computational time for this work was provided by ETH Zurich and the Swiss National Supercomputing Center (CSCS) under Project mr22. The Calculations were performed using the Piz Daint cluster at CSCS and Euler cluster at ETH Zurich.
  
\end{acknowledgments}


\bibliographystyle{apsrev4-1}

\bibliography{Biblio}

\end{document}